# SQUID MAGNETOMETRY FOR CANCER SCREENING
# A FEASIBILITY STUDY

G. G. Kenning[1], R. Rodriguez[1], V.S. Zotev[1], A. Moslemi[2], S. Wilson[2], L. Hawel[3], C. Byus[3], and J. S. Kovach[4]

[1]*Dept. of Physics, University of California, Riverside*
*Riverside, CA 92521*

[2]*Dept. of Biochemistry, University of California, Riverside*
*Riverside, CA 92521*

[3] *Division of Biomedical Sciences, University of California, Riverside*
*Riverside, CA 92521*

[4]*Long Island Cancer Center, Stony Brook University Stony Brook, NY 11794*

## Abstract

The recent demonstration that nanoparticles associated with various biological molecules and pharmacological agents can be administered systemically to humans, without toxicity from the particles, has opened a new era in the targeting of such particles to specific tissues in the body for the imaging and therapy of disease. The majority of particles used for this purpose contain iron and are detected in the body by magnetic resonance imaging. We believe a superconducting quantum interference device (SQUID) could provide quantitative and spatial information relevant to localization of superparamagnetic nanoparticles directed to a specific cell target in vivo. We envision a scanning system consisting of a DC induction field, a transport device, and an array of planar first order gradiometer coils coupled to DC SQUID amplifiers. We performed a set of computer simulations using experimentally determined values for concentrations of paramagnetic particles achievable in specific tissues of the mouse in vivo and concentrations of particles linked to monoclonal antibodies specific to antigens of two human cancer cell lines in vitro. An instrument to target distance of 10 centimeters was selected so that for an average adult scanning both the anterior and posterior surfaces could provide coverage of most of the body. The simulations demonstrate the feasibility of SQUID magnetometry for monitoring achievable concentrations of superparamagnetic particles in vivo and raise the possibility of using this approach to detect and localize collections of abnormal cells targeted by such particles.

## *Introduction*

Diagnosis of cancers at an early stage of development allowing complete surgical removal remains the most effective means of curing these diseases despite intensive efforts for half a century to develop curative pharmacologic regimens. Technological advances in imaging of abnormal structural features in organs has led to more effective methods for population screening for cancers but the successes are vastly outnumbered by the failures. For example, the overall cure rate for breast cancer in the United States is about 85% because there are effective if imperfect methods for early detection e.g. mammography and physical examination. However, for a tumor such as ovarian cancer, which produces few symptoms until it has spread regionally and for which there is no satisfactory method for population screening, the overall cure rate is about 35%. Detected at an early stage, usually by chance, ovarian cancer is almost 100 % curable.

Thus, much cancer research has concentrated on developing better methods for detecting the presence of the disease at an early stage with a low cost non-invasive technique suitable for population screening[1]. The most widely used methods for cancer imaging include plain x-ray with or without contrast medium, x-ray computed tomography (CT), magnetic resonance imaging (MRI), ultrasound, and positron emission tomography (PET). The most precise imaging techniques, MRI, PET, and CT are expensive and prohibitively so for large scale screening for cancer.

Over the past decade, there have been many attempts to develop non-toxic reagents that target abnormal collections of cells and which can be detected by a non-invasive procedure. One of the most promising approaches is the use of superparamagnetic nanoparticles (spnps) that are either actively concentrated into tissue such as lymph nodes, often the first site of cancer spread, or which are directed selectively to specific cell types by targeting molecules bound to the surface of the particles.

Superparamagnetic nanoparticles, consisting usually of iron oxide cores surrounded by dextran, are remarkably non-toxic in vivo in animals and humans. Thus, spnps linked to monoclonal antibodies with specificity to unique or overexpressed surface antigens have been used extensively to purify bone marrow stem cells constituting 0.1% or less of the total cell population, from unprocessed samples of human bone marrow. The antibodies cling to stem cell allowing their isolation from the mass of cellular material by magnetic column chromatography. The particles neither impair the ability of these stem cells to reconstitute the marrow of a leukemia patient nor do the particles appear to have any other toxic effects in humans.

Linking peptide sequences of the Tat protein of human immunodeficiency virus-type 1 (HIV-1) particles can enhance the uptake of spnps into specific cell types. Using this approach, Lewin et al[2] showed concentrations of 10 to 30 pg of superparamagnetic iron (0.5-2x$10^7$ nanoparticles/cell) could be incorporated into several types of human and mouse cells. The particles did not affect cell viability and human bone marrow progenitor cells ($CD34^+$ ) homed to the bone marrow in immunodeficient mice. Individual cells loaded with spnps could be detected as signal voids in MRI images and the presence of spnps in specific organs eg, liver, were readily detected in vivo by a reduction in MRI signal intensity, in the environment of the particles.

A striking demonstration of the safety and potential value of spnps in detection of prostate cancer metastatic to pelvic lymph nodes was reported recently by Harisinghani et al[3] . They administered spnps at a dose of 2.6 mg per kilogram of body weight to eighty patients with various clinical stages of prostate cancer. The particles concentrate in lymph nodes as one of a few preferential sites of tissue localization. By performing MRI before and after particle administration, 90.5 % of nodes found to have metastatic cancer (documented histologically after surgical resection or biopsy) after spnps compared to only 35.4 percent detection by standard MRI (before spnps). The basis for enhanced detection was distortion of the plain MRI image by the non-uniform suppression of the intensity of the T2 signal by spnps irregularly distributed in lymph nodes infiltrated by cancer cells.

A more direct method to capitalize on the magnetic susceptibility of spnps for cancer therapy is in development. Iron oxide particles bound to cancer chemotherapeutic agents are injected intraaterially and drawn through the walls of capillaries into the site of a cancer by an external magnet placed over the target site[4]. A clinical Phase I/II trial was recently completed in which the anti-cancer drug doxorubicin absorbed onto microparticles was magnetically concentrated in the liver of patients with unresectable hepatocellular cancer. Concentrations of the drug at the target were not measured in patients but localization of the drug in the liver, caused by the by the magnet field, was assumed because of little doxorubicin in the systemic circulation. Evaluation of the therapeutic benefit of this approach has been approved for evaluation in a multinational clinical study (see web site for FeRx, Inc.:www.FeRx.com).

Spnps offer many possibilities for specific cellular, disease and organ targeting for imaging and, potentially to achieve higher concentrations of molecules at sites of disease than are obtainable following systemic administration. We believe the low toxicity of spnps in vivo in humans and the ability to target specific tissues, including cancers in vivo, make them their inducible magnetic fields attractive as a new modality for cancer screening. The challenge is to create a detector sufficiently sensitive to rapidly acquire data on foci of increased magnetic signal in spatially defined volumes, which reveal the presence, and location of disease.

The objective of this study was to determine the physical parameters involved in detecting and localizing the magnetic field created by superparamagnetic particles localized on the tumor through phagocytosis or attached to tumor antigen-specific monoclonal antibodies targeted to occult human cancers  Over the last two decades dextran coated magnetic nanoparticles have found a variety of applications in the biological and medical sciences. Molday and Mackenzie[5] first describe the use of dextran-coated particles with a 15 nm iron oxide core coupled to antibodies and other ligand for cell separation in the laboratory. More recently a variety of smaller particles MIONS (Monocrystalline Iron Oxide Nanocolloid) [6], PION (Polycrystalline Iron Oxide Nanocolloid)[7], LCDIO (Long Circulating Dextran-coated Iron Oxide)[8] and USPIO (Ultra Small Superparamagnetic Iron Oxide)[9] have found application as contrast agents in MRI studies. These particles have been introduced in vivo and are found to be non-toxic. Maximum localization on a specifically targeted region (tumors, liver, lungs, lymph nodes) occur approximately 24 hours after injection and typically the majority of nanoparticles have passed out of the body within a week.

There are approximately $10^9$ cells in one gram of tissue.  If these cells formed a sphere, there would be about $10^6$ cells comprising the surface.  The recent demonstration by Lewin et al[2] that , in vitro , biologic transporter molecules such as certain peptide sequences of HIV-1 attached to spnps can achieve intracellular concentrations of iron of 10 to 30 pg of iron per cell or 10 to 30 ug per $10^6$ cells, an amount for paramagnetic iron which should be detectable by induction of a local magnetic signal arising from the tumor volume.  Such concentrations have not yet been achieved in vivo but as methods of targeting cancer cells in vivo improve, achieving such concentrations of spnps in early stage cancer seems within the realm of possibility.  Even current methods of targeting result in concentrations of spnps in experimental tumors that we believe should be detectable by virtue of their magnetic susceptibility.  Thus Moore et al [8] achieved concentrations between 11.9ng and 118ng of iron per million tumor cells growing as a mass in a rodent model after systemic injection of LCDIO.  We have modeled the ability of ultrasensitve SQUIDS to detect the magnetic fields that would be generated by these concentrations of iron.

Ultra-sensitive SQUID (Superconducting Quantum Interference Device) amplifiers have been used for imaging the magnetic fields produced by electrical activities in the brain and heart. Naturally occurring biomagnetic signals in the body are very small it was not until the advent of SQUID technology that practical measurements of these signals have been attainable. Since the advent of SQUID technology much effort has gone into imaging and mapping the biomagnetic signals produced by electromagnetic signals in the brain and heart. SQUID helmets with large numbers of array elements[10] (>100) have been produced to measure and map brain signals. Magnetocardiography (MCG)[11] and Magneto-Encephalography (MEG)[10] signals are time varying and produce changes in the magnetic flux outside the body. Since the inherent sensitivity (noise limit) is about 2 x $10^{-14}$T $/\sqrt{Hz}$ [12], operational bandwidths, we are considering (.1-40 Hz), allow for sensitivities as low as approximately $1.5x10^{-13}$ Tesla. System design and the ambient fields in the vicinity of the probe determine noise levels. Using gradiometer configuration pick-up coils signal levels on the order of $2 \times 10^{-15}$T /cm are readily achievable with good system design[13] in an unscreened relatively low noise environment. A diagram, which summarizes the sensitivities of some of the more common magnetometry techniques as a function of frequency, is Figure12.1 in **Bioelectromagnetism**[14].

SQUID scanner systems have been developed for use in Non-Destructive Evaluation[15] (NDE) of materials. Limitations in the cost effectiveness and penetration depths have hindered the techniques commercial introduction and most SQUID NDE evaluation still takes place in university laboratories. SQUID scanners using an ac field[16] have been used to determine the magnetic susceptibility of human organs in the body. While this type of system could potentially be used for the application we are describing, the ac field introduces a range of issues (electronic stability, balancing, eddy current noise etc.) that are for the most part avoided with our design. R. Ilmoniemi et al.[12] developed a SQUID scanner with many of the attributes of the system we are proposing. The main difference being that they were looking for a ferromagnetic inclusion (acupuncture needle) and hence their system has no inducing magnetic field. Other similar systems, used to measure iron stores in the liver have been described in the literature[17]

We investigate the limits of SQUID detection and localization on spnps at concentrations and at distances required for use of this technology as a rapid low cost method for screening for many types of cancers.

***Simulations***

A system for measuring the signature of a localized superparamagnetic particles within the body must have the following properties:

1) An induction field.
2) Method for inducing a magnetization induced flux change (ac field or motion).
3) Method for measuring a magnetization flux change.
4) Extreme stability
5) Extreme sensitivity

While the physical parameters to be measured are necessarily design dependent, we have chosen to simulate the system described below as it appears to be a straight forward and solves some of the problems inherent in the design of this type of measurement.

The initial simulations were performed to identify theoretical limitations and determine the physical parameters under which a SQUID coupled sensing device could obtain a signal from a superparamagnetic inclusion located in the body. To do this we developed a brute force three-dimensional simulation of the body, superparamagnetic inclusion, applied magnetic field, and sensing coils. By necessity a scanner design must be chosen and a particular protocol simulated. We have chosen a scanner, which incorporates a DC superconducting induction field (low noise and highly stable) and uses motion of the patient to produce a flux change in the pick up coils. This design is simple, cost effective and flexible.

The magnetic field was modeled as a racetrack geometry extending across the width of the body (Figure 1). Neglecting end effects (extending the magnet well past the width of the body), the magnetic field could then be modeled as a contribution from two wires separated by 14cm, each wire located 7cm on opposite sides of the scan point. To minimize ambient noise and the background signal, the pickup coils were modeled in a planar first-order gradiometer configuration[18] with each of the counter wound pickup coils having an area of $1cm^2$ and each coil located along the length-axis with the center of the coils displaced 2.5 cm on either side of the scan point, Figure 1. The magnetic field signal due to the inclusion and/or a background cube is calculated at a point at the center of each coil.

The phantom torso (body) in our simulations was modeled as an ellipsoid filled with water. This approximation was used to determine the maximum absolute contribution of a large slowly varying diamagnetic background (of approximate torso dimensions) with rapidly varying edges. There are clearly limitations on the information that can be obtained from this simple model but the model is useful for determining some of the limitations of the technique.

A three dimensional rectangle was formed with dimensions of length 120cm, width 20cm and thickness 20cm, Figure 2. The dimensions of the rectangle were then divided into $mm^3$ cubes and the magnetization from each cube contributes to the measured signal. The torso was modeled in the rectangular box as an elongated ellipsoid with the same maximum dimensions as the rectangle. During the summing of the magnetic field contributions of the magnetization, cubes located outside the ellipsoid produced no contribution to the field at the sensing coils. Cubes located inside the ellipsoid had a diamagnetic response to the applied field and produce a corresponding contribution to the field at the SQUID sensing coil. The Molar Diamagnetic susceptibility of water is $-13 \times 10^{-6}$ (emu/mole). The tumor was modeled as a paramagnetic inclusion and could be located at various positions within the ellipsoid. The tumor with magnetic particles was represented as the contribution from 1mm square cube located within a 1cm diameter sphere. These “tumor” cubes had both a diamagnetic contribution (body) and a paramagnetic contribution due to the magnetic nanoparticles. Experimentally determined values for the paramagnetic contributions will be described in the next section.

The magnetic field located at a distance r from a solenoid wire has a magnitude equal to B(r)=C/r where C is a constant determined by the magnetic field at 1 cm. This field was taken to be 3500 G, producing a total magnetic field of 1kG at the central scan point. The field is a vector quantity radiating tangentially from a circle or radius r centered on the wire. The field due to the second wire circulates in the opposite direction giving a significant cancellation of the x components near the vertical line passing through the scan point. The y-components of the magnetic fields located near the same vertical line add, producing strong vertical polarization of the diamagnetic ellipsoid and the magnetically enhanced tumor. By knowing the direction of the magnetic field and magnetic susceptibility at any point in the matrix we can calculate the magnetization vector. Treating the magnetization of a $mm^3$ cube as a magnetic dipole we can calculate the magnetic field produced by the sample magnetization at the position of the pick up coils.

The scan covers the upper positive quadrant of the three dimensional rectangle and hence the upper quadrant of the ellipsoid. The rectangle and ellipsoid were shifted by 10cm from the X-Y plane to allow for a scan across the full width of the body. In general the pickup coils were placed 1 cm from the top of the ellipsoid. A scan at any one scan point includes a scan volume of 140 mm along the length, 200mm along the width and 100 mm of thickness and the magnetic field is calculated at each of the two counterwound pickup coils for each of the ten scan elements. The 140 mm finite length of the scan volume has the effect of clipping the generated signal at distances greater then or equal to 7 cm from the tumor along the length axis. As the scanner is moved the length of the body 600 of these scan volumes are included in the total scan. We estimate that the total 11 SQUID simulation includes approximately $1x10^{11}$ calculations and takes approximately 9.5 hours on a 960MHz Pentium III PC. Even so the $mm^3$ grain size appears as rapid jumps in the torso contribution as the $mm^3$ grains are limited by the smooth ellipsoidal function and in the tumor contribution as the $mm^3$ grains are limited by a smooth spherical function.

## ***Experimental Determination of Simulation Parameters***

### Magnetic Nanoparticles

Two types of targeted nanoparticle contributions to the paramagnetic moment of the tumors were considered. The first targeted nanoparticle we will consider is used in MRI contrast imaging. The paramagnetic contribution was calculated using the parameters given in Shen et al.[6], who studied the behavior of magnetic nanoparticles uptake by mouse brain tumors as a contrast agent for MRI. Tumor cell uptake of LCDIO, in a rodent model, was found to be between 11.9ng and 118ng of iron per million cells[8]. In this study the LCDIO data is being used to determine that the minimum signal and we therefore use a value of (12.5ng of iron)/(1 million tumor cells), at the bottom range of their reported results. Using the reasonable assumption that there are approximately $1x10^6$ cells/$mm^3$, we estimate that a 1 $cm^3$ tumor to contain 12.5μg of Fe. From the graph of the magnetization vs. field in Shen et al. we observed that the saturation magnetization occur for magnetic fields > 5 kG. In the field range we will be working, < 2 kG, the magnetization vs. field is approximately a straight line and from this observation we took the magnetic susceptibility for fields less then 2kG to be $2.2x10^{-2}$ emu/gm (Fe). Simulation results using these parameters are labeled LCDIO.

The second set of parameters which we used to simulate maximal targeting of the cancer tumor were obtained from in vitro targeting to specific lines of cancer cells with monoclonal antibody attached magnetic nanoparticles. These conjugates are produced commercially by Miltenyi Inc. for use in cell separation. Due to the high number of potential binding sites on the cancer cells, we chose to use Miltenyi's anti-cytokeratin monoclonal antibodies that were conjugated to magnetic nanoparticles. Simulation results using these parameters are labeled MACS.

LNCaP human prostate cancer cells and MCF-7 human breast cancer cells were grown at 37° C and 100% humidity in Dulbecco's modified Eagle's medium containing 10% fetal bovine serum. Both cell lines were routinely subpassaged to maintain them in log-growth phase. To label the LNCaP and MCF-7 cells with the magnetic anti-cytokeratin antibodies, the cells were harvested by quick trypsinization, and pelleted in a clinical centrifuge. Following a short wash with phosphate buffered saline solution, the cells were again pelleted and allowed to dry briefly. The antibodies were then bound to the LNCaP and MCF-7 cancer cells as per the manufacturer's protocol. Following antibody binding, the cells were counted with a hemocytometer and aliquoted into microfuge tubes for SQUID analysis.

To determine the physical parameters associated with using the MACS particles in the simulation we performed a series of experiments using a Quantum Design SQUID magnetometer. Miltenyi literature describes their particles as polysaccharide coated iron oxide nanoparticles approximately 50 nm in diameter. To obtain an estimation of the mean size of the iron oxide component of the nanoparticles we have measured the field cooled and zero field cooled magnetization as a function of temperature (Figure 3a). The blocking temperature of iron oxide nanoparticles is well-known as a function of temperature and from the observed blocking temperature we estimate that the iron oxide particles in the Miltenyi microbeads have a mean in diameter of approximately 6-10 nm. We do not expect that the distribution contains significant number of particles greater then 10 nm since this would produce of remanence at temperatures greater than the measured blocking temperature. To determine if the nanoparticles are truly superparamagnetic or if they have some remnant moment we measured in the in-phase and out-of-phase magnetic susceptibility of the particles, between 1 Hz and 1000 Hz, at 310 K (Figure 3b). it can easily be seen that these particles are paramagnetic. There is a large approximately constant component of the magnetic susceptibility and the out-of-phase component is very small over the whole frequency range in but does increase as a function of frequency.

To determine the magnetization values to be used in the simulation we measured the magnetization as a function of magnetic field for the particles targeted to the LNCaP and MCF-7 cancer cells at 310 K (Figure 4). One significant difference between the MACS particles and the LCDIO particles is that the MACS particles require a much lower magnetic field to obtain saturation magnetization. The data was fit to a Langevin function for a collection paramagnetic particles. From these fits the magnetization/particle was determined and found to be consistent with what would be expected for 8-10 nm sized $Fe_3O_4$ nanoparticle. The breast cancer cells were each determined to have picked up approximately 3.3 x $10^5$ MACS particles. The prostate cancer cells had binding values approximately one of order of magnitude less then that obtained for the breast cancer cells. We fit a $9^{th}$ order polynomial function to the magnetization (M) vs. magnetic field (H) curve and used the function M(H) in the simulations. At each cube of tumor cell, in the simulation, we first determine the value and direction of the magnetic field. Then using the M(H) function we determine the magnetization vector for that cube.

Based on the concentration of LCDIO nanoparticles achievable in tumor cells in vivo in a rodent model, we determined what we consider to be the minimum values required for detection of a useful signal in a screening system. The MACS data provide the maximum possible targeting of a specific monoclonal antibody to two types of cancer cell lines. The MACS data provide a more idealized situation but one which points to the possibilities inherent in this technique.

## Results

We first determined the feasibility of detecting the magnetic field generated by the maximum number of MACS expected to be localized in tumor with a volume of one cubic centimeter. To address this question, we need to consider the limits of SQUID technology and the effects of ambient noise. Our initial approach is to model detection of different strengths of generated signal assuming zero external noise and then to address the realities of background signal.

The absolute magnetic field generated by a 1 $cm^3$ tumor at various distances from the SQUID scanner is shown is Figure 5. In this range, the signals are above detection limits. For example, magnetocardiogram signals have been detected with SQUID scanners in the sub pico-Tesla range. The signal from the LCDIO particles at a depth of 11 cm is in the pico-Tesla range. The signal from the MACS particles attached to the human breast cancer, at this depth, is four orders of magnitude greater than those from the LCDIO particles.

Far field noise reductions of two to four orders of magnitude can be achieved with a well-balanced gradiometer configuration for the pick-up coils. Currently lithographic techniques used to pattern planar pickup coils can provide excellent balance. While the differential magnetic field detected by gradiometer configurations is smaller then the absolute magnetic field, this difference is more then compensated for by the reduction (cancellation) in ambient noise. Figure 6 is the scan, for an 11 SQUID scanner, neglecting the signal from the body. The spatial distribution of the differential scan signal at the various pickup coils, generated from a signal a depth of 5cm from the central pickup coil, is plotted. As the scanner moves across the length of the scan rectangle, the pickup coil on the near side begins to pickup the signal. The signal reaches a maximum close to the point where the pickup coil is positioned vertically over the tumor. The signal then goes to zero when the scan point is directly over the tumor and becomes negative as the other counterwound coil passes over the tumor.

We believe that the pickup coils must to be able to sense at a minimum of 10 cm into the body. For a maximum body thickness of 20 cm (for a patient lying on a flat table), the anterior and posterior surfaces would be scanned to include the entire anatomy. Allowing for 1cm clearance between the pickup coils and the body surface, the minimum scanning distance is 11 cm. The raw signal for a tumor located 11 cm from the pickup coils is shown if Figure 7 and the maximum differential signals (signal from the pickup coil) from the tumor as a function of the vertical distance of the tumor from the scan point of the detector (for both LCDIO and MACS targeted MCF-7 human breast cancer cells) are shown in Figure 8.

The magnitude of the signal for the LCDIO particles exceeds the resolution of the technique in an unscreened environment. However, the signal to noise ratio is probably not adequate for realistic and reproducible detection with an expected ambient noise of about $10^{-12}$ Tesla. The LCDIO particles represent a minimum signal that has been achieved in vivo. The targeted MACS produce a signal four orders of magnitude greater then that produced by LCDIO particles and are well within the detection limits of a SQUID scanner. The MACS particles represent a maximum signal for cell targeting and as such provide an upper bound for targeting a 1 $cm^3$ tumor with this particular size and type of magnetic nanoparticle.

We simulated the effects of the diamagnetic background of the body. A tumor located in an ellipsoid of the susceptibility of water and positioned 11 cm from the SQUID sensor is illustrated in Figure 9. The LCDIO value at 6 cm results in a signal approximately equal to the maximum signal of the diamagnetic background and can therefore be resolved. It should be noted that the background has a noise associated with it, which appears as ripples. These “ripples” are an artifact of the algorithm, as previously described, but provide an artificially implemented noise level at approximately 0.1 nano-Tesla. While signals in this sub pico-Tesla regime have been measured in scanners, it is likely that the addition of the superconducting solenoid with a DC magnetic field of approximately 0.1 Tesla will significantly increase the noise level. Unlike a Helholtz coil where the vast majority of noise is induced through the electronics, a superconducting solenoid in its persistent mode is very stable and introduces very little noise. We expect the main source of noise to be from vibrations of surrounding materials in this magnetic field. We believe a scanner, operating in the described magnetic fields, can be constructed with an intrinsic noise level of ~0 .1 nano-Tesla. We selected this level of noise to begin to set limits on this technique. This artificial noise is more then an order of magnitude larger then the signal of an LCDIO tumor at 11 cm making detection impossible. To resolve a 1 $cm^3$ tumor at 11 cm, at this noise level, requires approximately two orders of magnitude more signal strength. Such signals are likely to be achievable as in vivo targeting of spnps evolve.

To consider background signal, we investigated the effects of a uniform background signal upon detection of a tumor located in a large organ. Moore et al.[8] considered this issue with regard to 9L gliosarcoma brain tumors labeled in vivo in a rodent model.. In this study, uptake by the brain tumor was approximately 0.11% of injected dose and the surrounding normal brain tissue had about on tenth of the tumor concentration of LCDIO. The effect of this type of background signal was investigated in Figure 11. We modeled a 1.0 $cm^3$ tumor located in an organ at 5 cm from the scanner. The organ was modeled as a rectangular box with thickness 2.5 cm on either side of the tumor, width 5cm on either side of the tumor and length 5 cm on either side of the tumor in the direction of the scan length. The organ was given an Iron Oxide concentration of 1% of the tumor concentration. The diamagnetic background of the body was included. The scan in Figure 11 shows that a 1% organ background produces a signal comparable to the signal due to the tumor. At 1% the signal due to the tumor is still resolvable. Simulations were also done with a 10% background tumor. In these simulations the background signal is orders of magnitude greater then the tumor signal. Subtraction of this background signal is problematic and sets limits on contributions due to surrounding tissue. On the other hand for intravenously administered targeted Iron Oxide particles conjugated with monoclonal antibodies it was found that tissue surrounding the tumor had “modest” uptake but no evidence of the presence of monoclonal antibodies. This suggests that well constructed magnetic label-target specific vector conjugate will provide the best system for maximizing tumor uptake while minimizing background.

The body scans in the plane of the scanner, the XY plane, show that localization of the signal is a function of the number of pickup coils comprising the scanner. There are several attributes of the signal that may be used to determine depth in the body. One well-known attribute is that the peak of the differential signal changes in position as a function of depth (Figure 1a). Another perhaps more useful way of determining depth is suggested by the form of the magnetization vs. the magnetic field curve in Figure 4. The deeper the tumor in the body, i.e. from the greater the distance to the scanner, the smaller the magnetic field.

When the tumor signal is in the XY plane, a series of shorter length scans can be performed with increasing magnetic field. This type of field scan, using the signal obtained from MACS attached to MCF-7A cells, is illustrated in Figure 11b. The data as a function of magnetic field can be fit reasonably well by a power law and the exponent of the power law changes significantly as a function of tumor depth.. Therefore, it is possible to expand the signal as a function of the power law's and determine the signal has a function of depth.

## Discussion

In a practical embodiment of the proposed scanner, the magnetic field solenoid and scan elements would be located beneath the patient facing upward. The patient would be moved over the scanner on a flat conveyor system without contact with scanner. Thus, the scanner is isolated from vibrations associated with patient movement. The magnetic field solenoid and pickup coil loops must be rigidly connected to each other. In magnetic fields of a few kG, vibration or motion of the pickup coils will produce noise levels greater than the sample signals.

The SQUID sensor is located approximately 30 cm from the pickup coils in a Nb enclosure. The magnetic field should not affect the SQUID. Before a scan, the SQUIDs can be reset to lock the trapped flux due to the superconducting solenoid in the pickup coil loops. Variations of magnetic field are measured with respect to this constant amount of trap magnetic flux. At reasonable transport rates of 5 to 10 cm per second so that a body scan would be completed in 20 to 40 seconds. This rapid scan rate allows multiple scans to be averaged, reducing the signal to noise ratio. SQUID drift would be measured measured before a scan and compensated for over the duration of the scan.

While SQUID magnetometers have intrinsically high sensitivity, their behavior is usually limited by the ambient noise of the environment. In the measurement mode considering, the SQUID effectively integrates its signal over a large frequency range. Although, noise can occur over a large range of frequencies. High pass filtering (> 1 Hz), by shorting the SQUID terminals with copper wire, can reduce the inherent low-frequency noise.. A low pass filter less then $10^4$ Hz will reduce noise from radio waves. However, geomagnetic noise, general low-frequency noise due to motion of cars, trucks, air-conditioning, human activity, etc., and line frequency noise associated higher harmonics, may require enclosing the scanner and transport mechanism in a room with electromagnetic and mu metal shielding. Specific filtering of the SQUID signal at the line frequency and higher harmonics may also be needed..

In terms of the physical parameters analyzed in this simple model, values of magnetization in realistic magnetic induction fields are measurable with SQUID technology. There are several variables that can be potentially modified to enhance the signals generated using this technique. It is clear that the 4 nm LCDIO particles, and uptake rates reported from MRI studies, provide enough signal for a 1 $cm^3$ tumor to be detected at 10 cm depth in the body. Even so the signal to noise ratio will be poor. Uptake of the LCDIO particles occurs through phagocytosis and as such they are not specifically targeted to the tumor cells. The LCDIO samples are also not specifically designed to maximize the signal in a SQUID scanner. Data from the MACS targeted samples offers an example of what maybe possible. The MACS data is four orders of magnitude greater then the LCDIO data and would provide sufficient signal to noise ratio for a 1 $cm^3$ tumor to be observed at distances greater then 12 cm from the scanner. In the body one would not expect this type of the maximal targeting that we have presented from in vitro combination of cells and targeted nanoparticles. One might expect targeting ratios of the few percent of possible antigen sites thereby dropping the signal by one to two orders of magnitude. One possible direction for enhancing signal is therefore to maximize nanoparticle to tumor binding. The main parameters available for enhancement of the signal are related to the characteristics of the nanoparticles. Nanoparticles moments, above the blocking temperature, increase as a function of the radius of the nanoparticles[19]. It could be argued that this increase goes approximately as the radius cubed. Increasing the size of the iron oxide nanoparticles from a radius of 5 nm to radius of 25 nm (the maximum size for single crystal iron oxide nanoparticles) would give a signal increase of approximately two orders of magnitude. This increase in size would also significantly raise the blocking temperature and increase resonance. There is a large amount of work being done to develop nanoparticles for a variety of applications. For this application ideal nanoparticles would have the following properties:

1) A blocking temperature less then 310K,
2) Maximized magnetic moment (as a function of size and constituents)
3) Non-toxic
4) Small enough to pass though the endothelio-reticular system.
5) Coated to evade the bodies defenses

We also analyzed a 1$cm^3$ tumor at 5cm within a simulated organ of thickness, width and length, 5cmx10cmx10cm consisting of a diamagnetic water concentration and a 1% of tumor concentration superparamagnetic contribution. The organ contribution was as large as the tumor signal suggesting the need for efficient targeting. One other way to reduce contributions from an organ would be to measure and then model the organ uptake and subtract the signal from the scan. Theoretically any tumor with signal greater than the minimum noise constraints could be identified.

In conclusion, magnetic field signals from a 1 $cm^3$ tumor loaded with in vivo determined values (LCDIO) of iron oxide nanoparticles produces significant enough signal to be measured with a DC SQUID scanner at depth of 11 cm. However when the expected level of noise for this particular type of scanner, is taken into account, the signal is not resolvable. It must be pointed out, that these contrast agents were not designed for this type of magnetization measurement and hence it is likely that the signal could be greatly improved. Nanoparticles targeted to cells in vitro (MACS) produce signals four orders of magnitude larger than those produced by LCDIO particles. At these signal levels tumors, at depths greater than 11 cm, can be detected by a SQUID scanner with signal to noise ratios greater then 100: 1. We believe that highly efficient tumor targeting and maximization of nanoparticle magnetic moment will be achieved in the near future making SQUID magnetometry an important modality for cancer screening.

The authors would like to thank Bob Fagley, Bob Kraus and Raymond Orbach for useful discussions.

## Figures:

**Figure 1. Illustration of hypothetical SQUID scanner modeled in simulations.**
Scanner is composed of ten planar first order gradiometrer pickup coils located to the interior of and on the same platform as a superconducting solenoid. The value of the magnetic field is the field generated at the midpoint of a set of coils of a pickup coil. This is also defined as the scan point. End effects of the solenoid are neglected.

**Figure 2: Computer Simulated Body, Tumor and Scan Geometry**
The scan covers the upper positive quadrant of the three dimensional rectangle and hence the upper quadrant of the ellipsoid. The rectangle and ellipsoid were shifted by 10cm from the X-Y plane for mathematical simplicity (all positions in this quadrant are positive definite). The pickup coils, located in the scanner are located 1 cm above the top of the rectangle. The magnetic field is in a racetrack configuration centered around the scanner platform.

**Figure 3: Magnetic Properties of MACS particles.**

a) The zero field cooled and field cooled magnetization of 50 microliters of MACS solution measured in the magnetic field of 100 G.
b) b) The in-phase and out-of-phase components of the ac susceptibility as a function of frequency of 50 microliters of MACS solution.

**Figure 4: Magnetization Vs Magnetic Field**

**a)** $10^6$ cells MFC-7a MACS conjugated Human Breast cancer cells,
**b)** $10^6$ cells LNCaP MACS conjugated Human Prostate cancer cells.

**Figure 5: Absolute magnetic field (single coil) at the scanner due to enhanced magnetization as a function of perpendicular distance from the scanner.** Distance to scanner varies from 2cm to 11 cm. Edge effects of signal are due to a finite scan volume width and accentuated by the log scale. The magnetic field applied was .1 T (1000 G) at the scan point.

**Figure 6: Line scan of magnetic field differential vs. Scan Distance MACS conjugated to MCF-7 cells**
No background. Tumor is located at x=10 cm (100mm) y=6 cm (5 cm depth from pickup coils) and z=10cm (center of the ellipsoid). Scan produced by 11 SQUID scanner.

**Figure 7: Signal From Tumor Located 11cm deathblow the SQUID Sensor**
Tumor is located at x=10 cm (100mm) . Scan produced by 11 SQUID scanner. Maximum differential signal is plotted as a function of the scan distance.

**Figure 8: Differential Magnetic Field Signal Maximum as a function of distance of 1 $cm^3$ tumor from the scanner.** The magnetic induction field applied was .1 T (1000 G) at the scan point.

**Figure 9: Differential magnetic field scan (output from first order pickup coil subtraction) as a function of scan distance.**

a) A 1$cm^3$ LCDIO tumor located 100 mm from the y-axis and located at a distance of 5 cm from the central scan point and a depth of 4 cm below the surface of the diamagnetic ellipsoidal background.

b) A 1$cm^3$ MCF-7 MACS targeted constructed tumor located 100 mm from the y-axis and located at a distance of 11 cm from the central scan point and a depth of 10 cm below the surface of the diamagnetic ellipsoidal background.

Simulated signal generated by 11 SQUID Scanners scanning the width of the ellipsoid. The magnetic induction field applied was .1 T (1000 G) at the scan point.

**Figure 10: Line Scan: Tumor at 10cm plus Virtual Organ Background plus Diamagnetic Background**

For ellipsoidal background. Tumor is located at x=10 cm (100mm) y=1cm (10 cm from pickup coils) and z=10cm (center of the ellipsoid). Scan produced by 11 SQUID scanner.

**Figure 11: Depth Dependence of Scanner Signals.**

a) The position of the peak in the in the Differential Magnetic Field signal from the tumor as a function of tumor depth.

b) Amplitude of the maximum Differential magnetic field signal from the tumor as a function of magnetic field for tumor is located and varying depth.

The magnetic induction field applied was .1 T (1000 G) at the scan point.

.

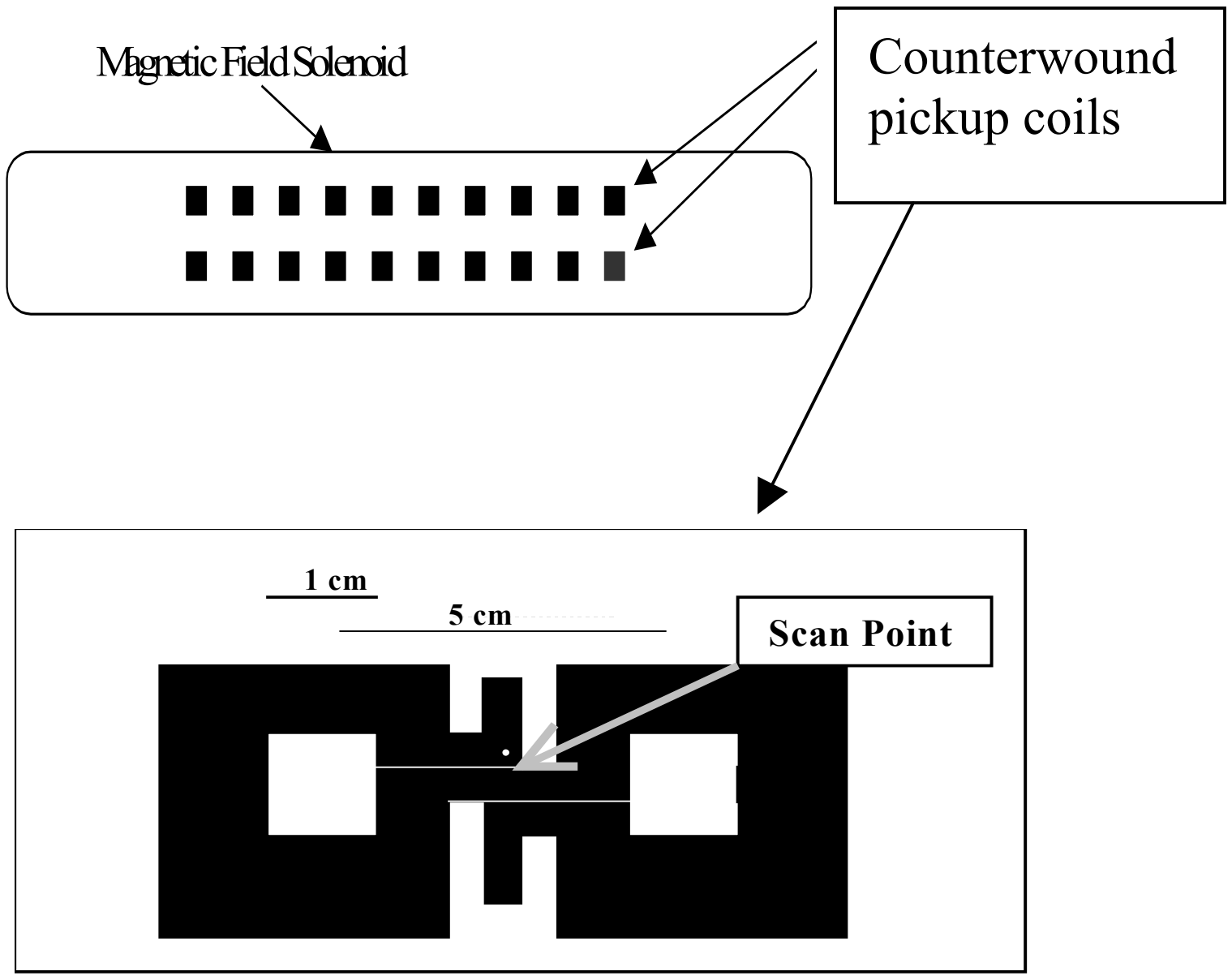

**Figure 1.**

.

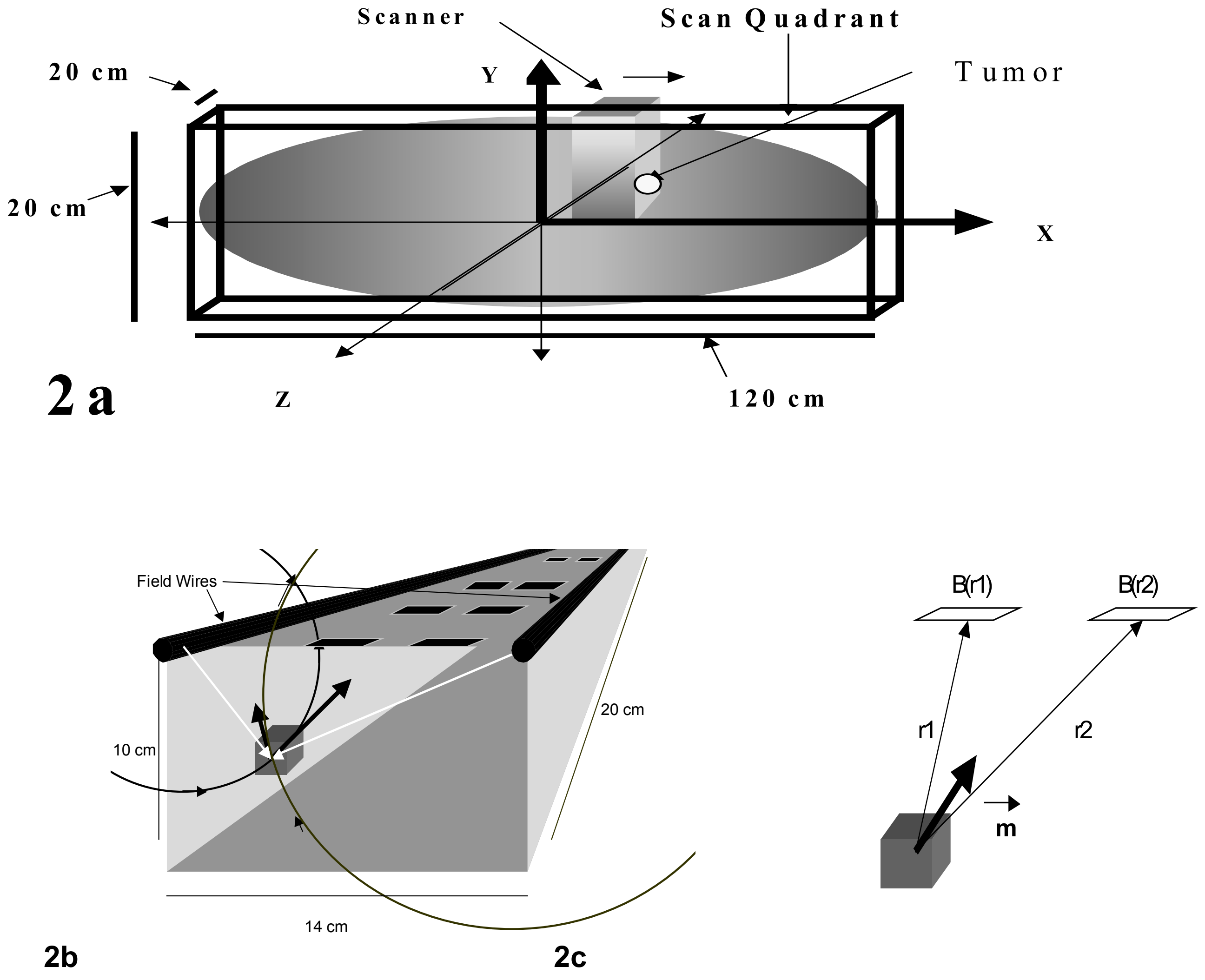

**Figure 2: Computer Simulated Body, Tumor and Scan Geometry**

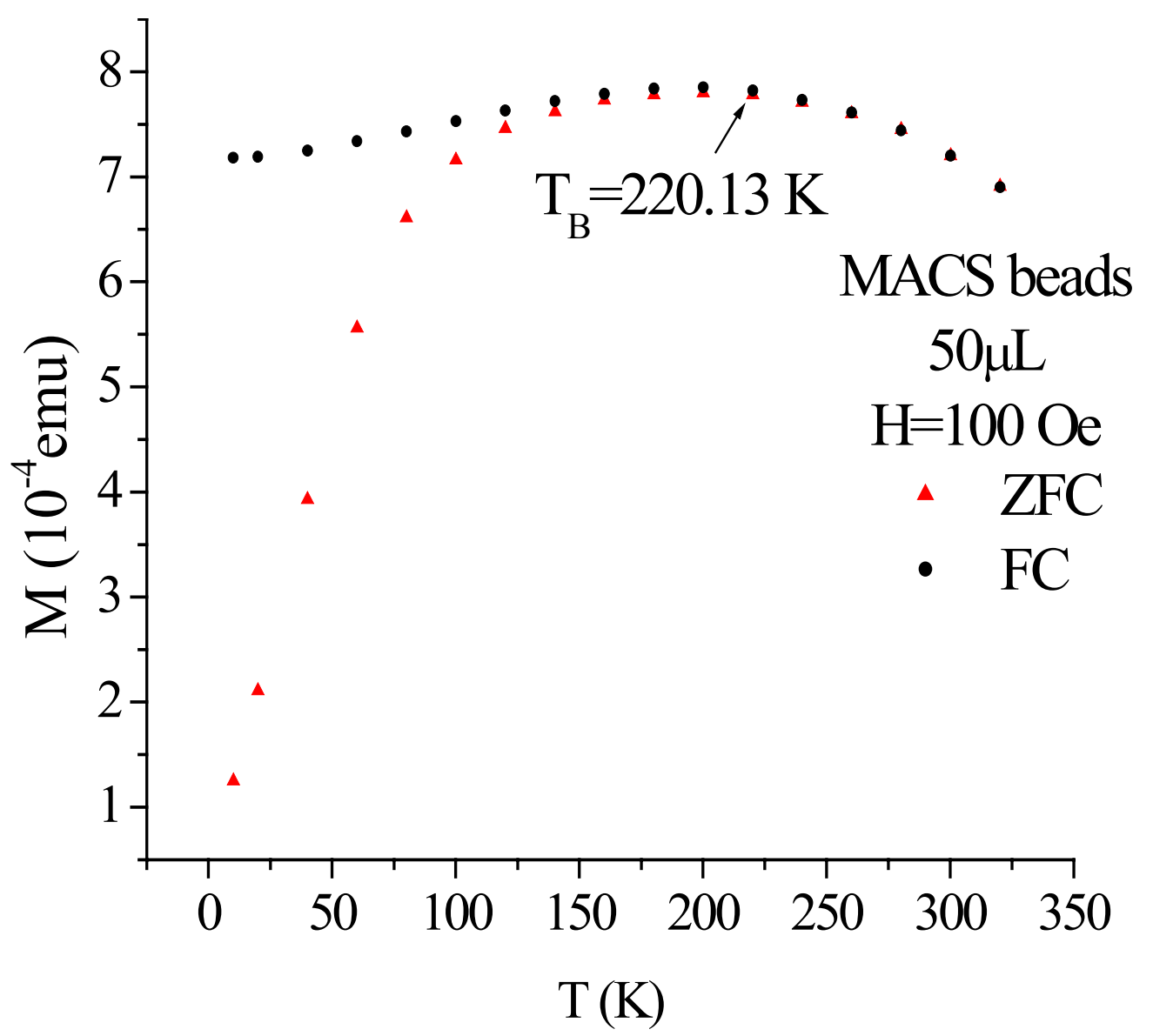

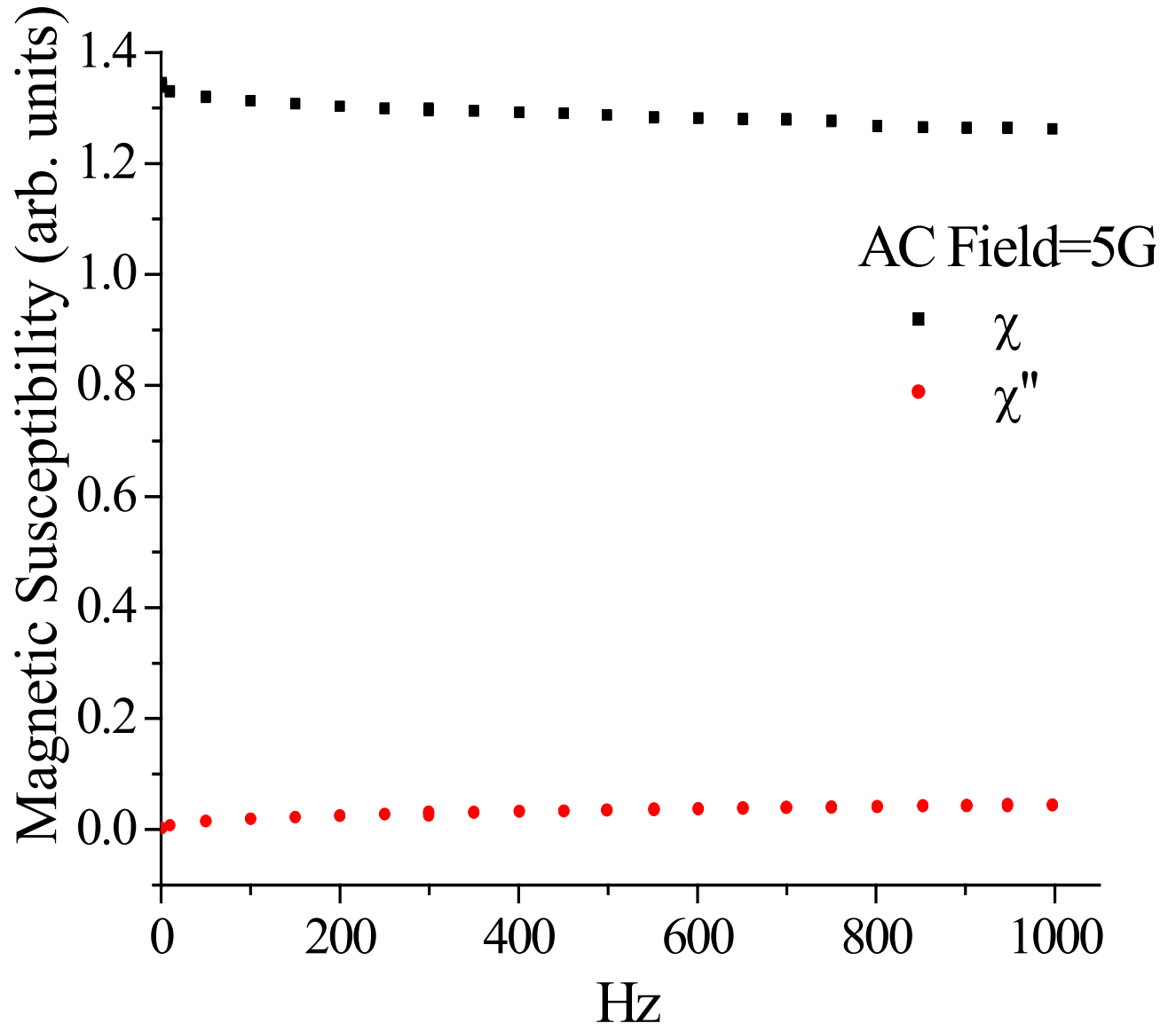

**Figure 3: Magnetic Properties of MACS particles.**

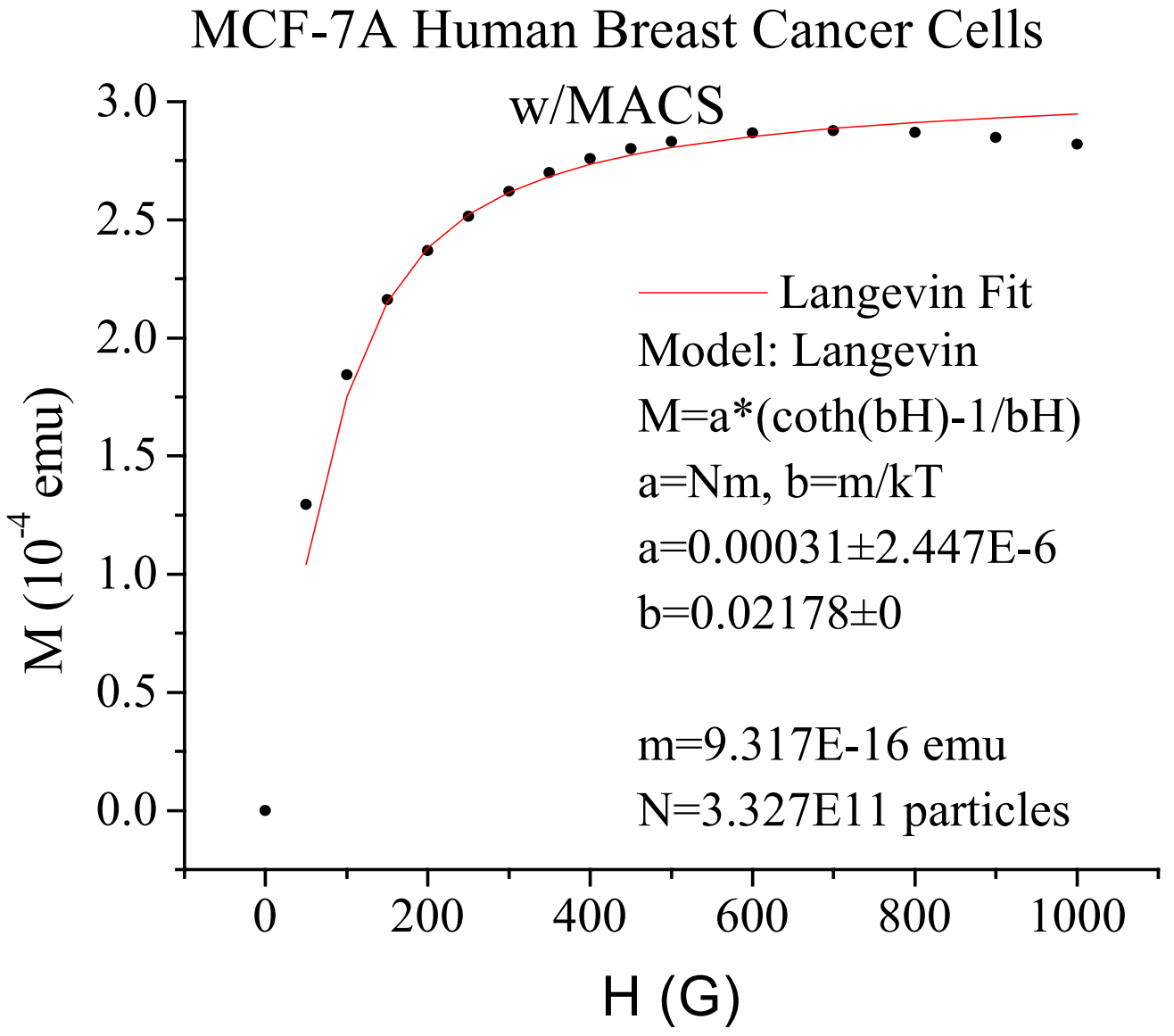

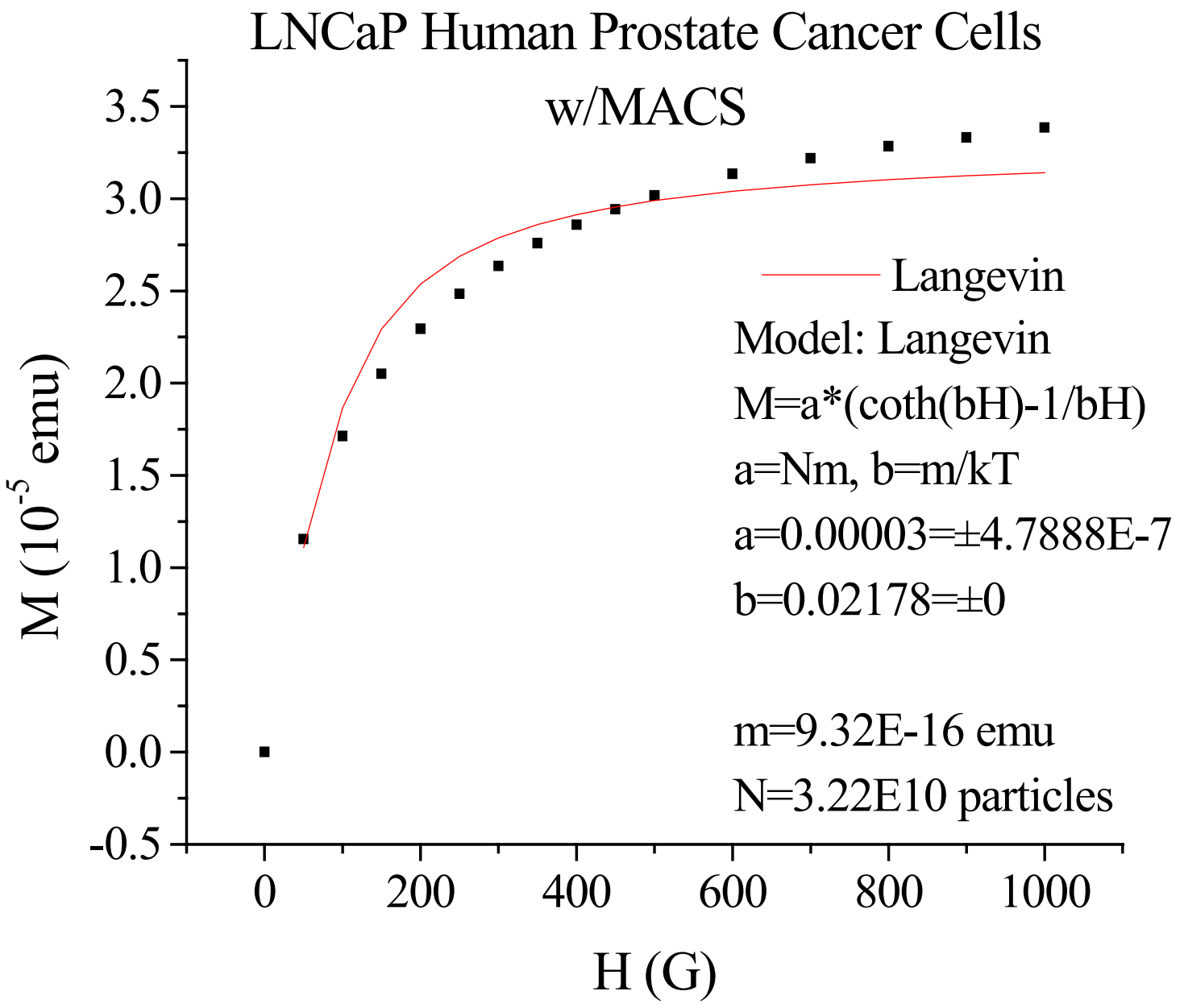

**Figure 4: Magnetization Vs Magnetic Field**

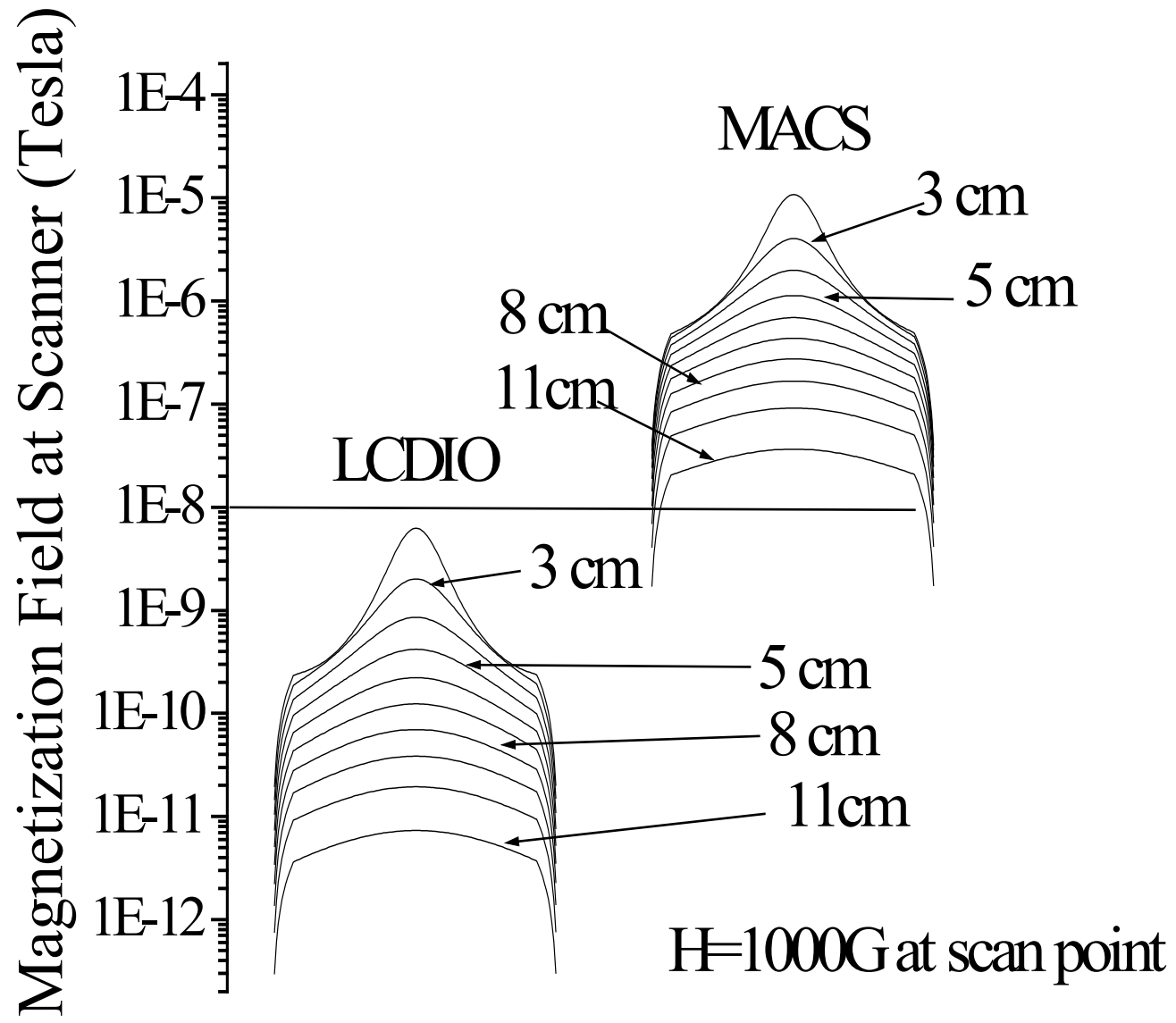

**Figure 5: Absolute magnetic field (single coil) at the scanner due to enhanced magnetization as a function of perpendicular distance from the scanner**

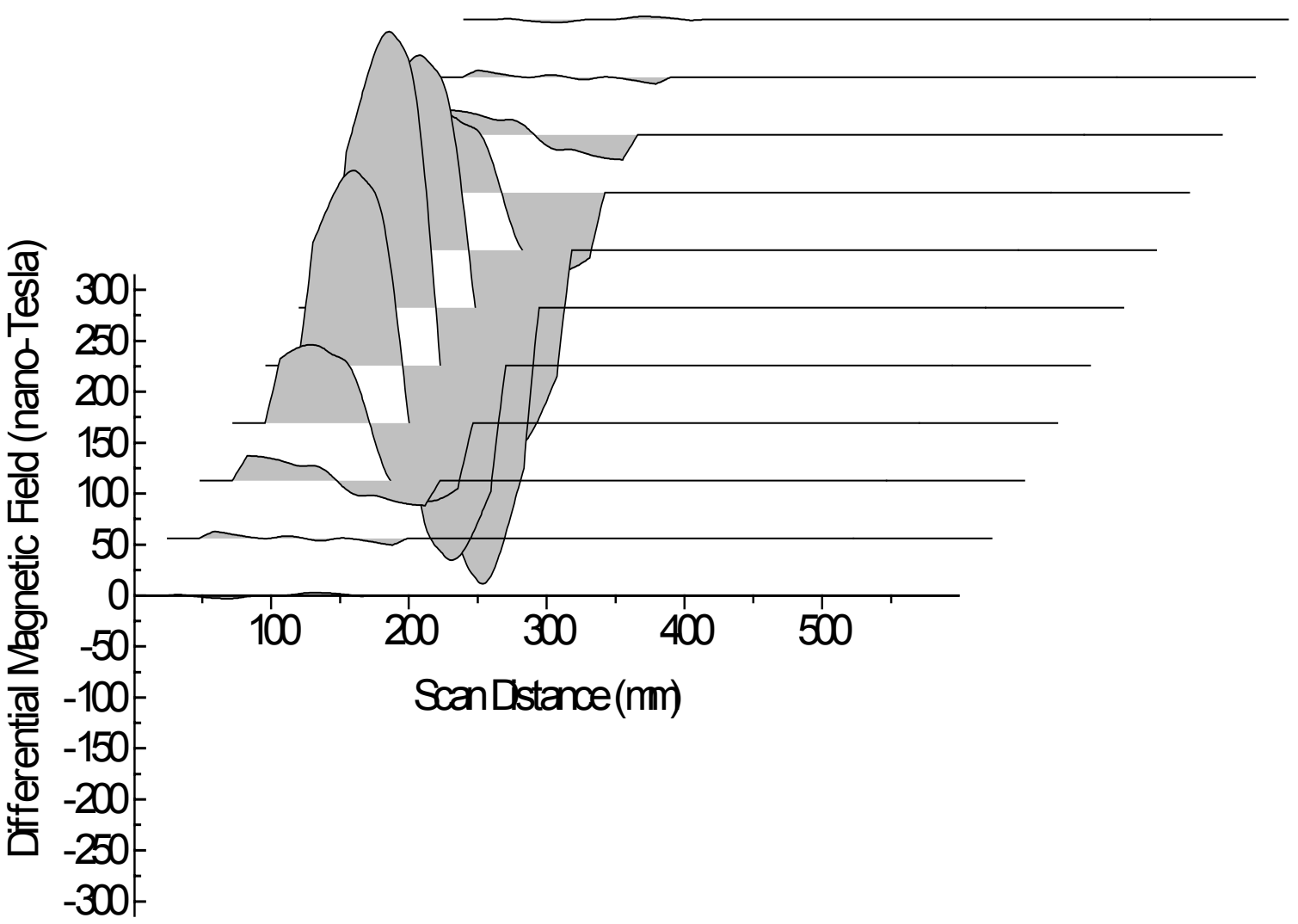

**Figure 6: Line scan of magnetic field differential vs. Scan Distance MACS conjugated to MCF-7 cells**

.

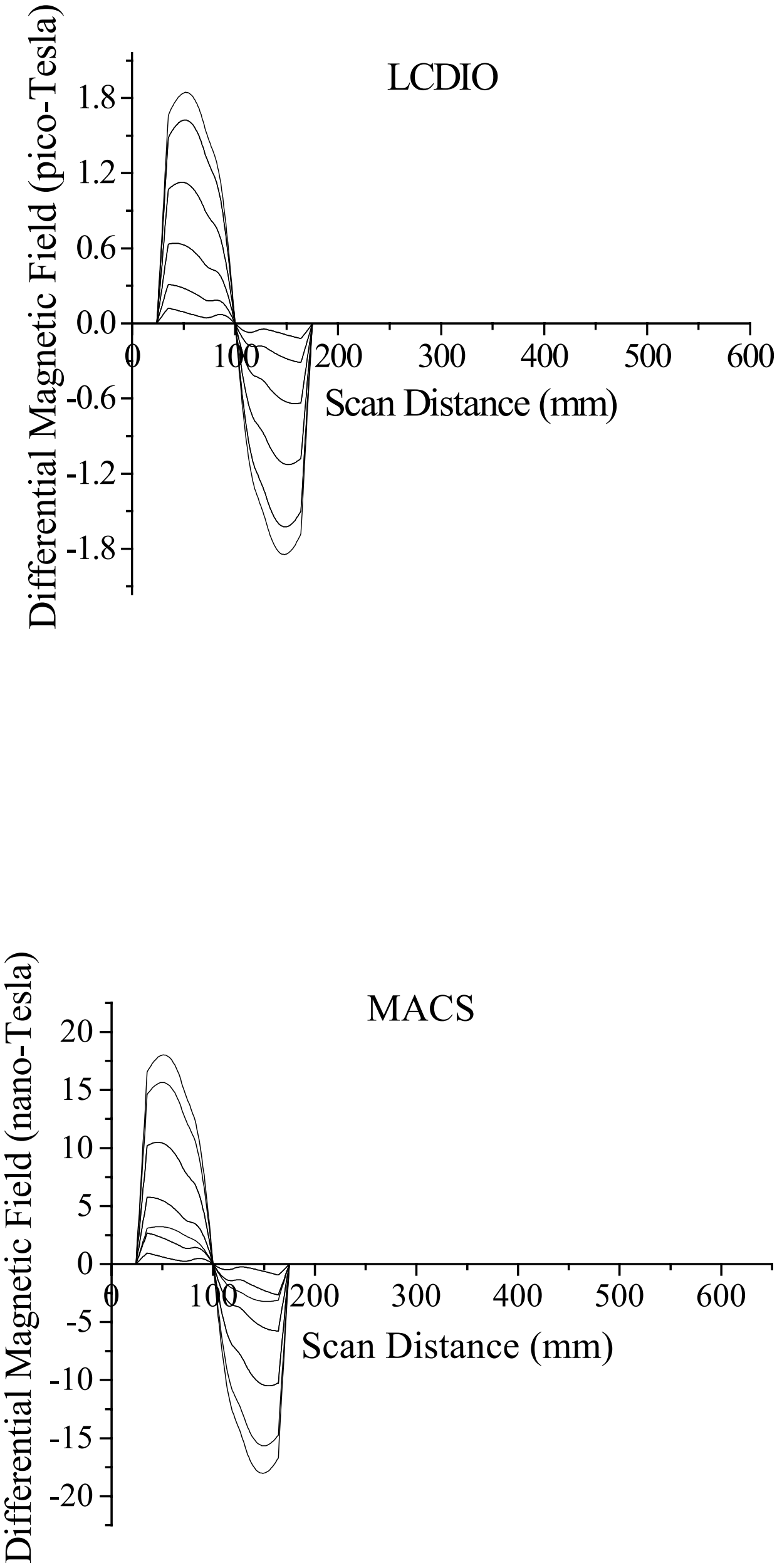

**Figure 7: Signal From Tumor Located 11cm deathblow the SQUID Sensor**

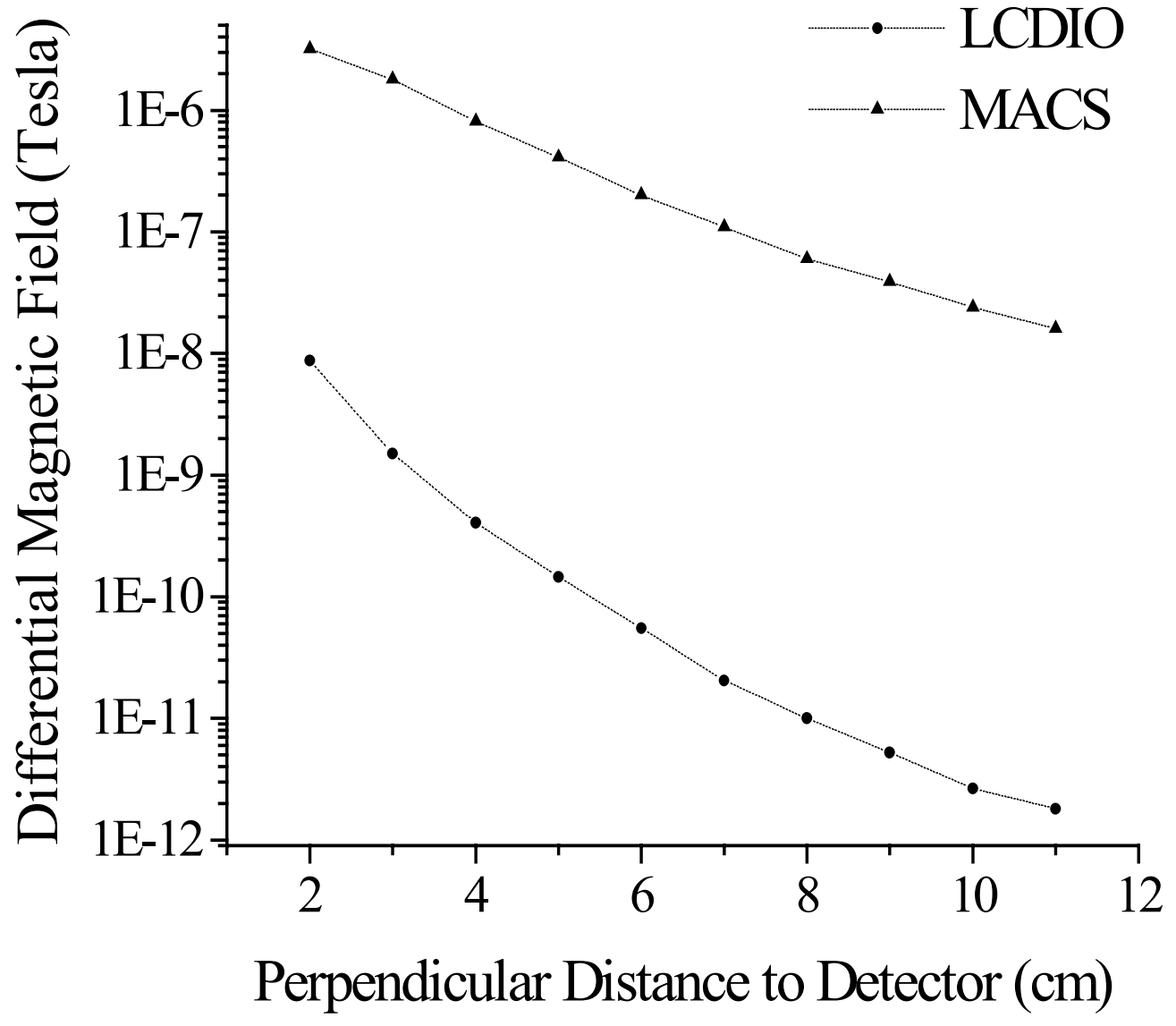

**Figure 8: Differential Magnetic Field Signal Maximum as a function of distance of 1 $cm^3$ tumor from the scanner.**

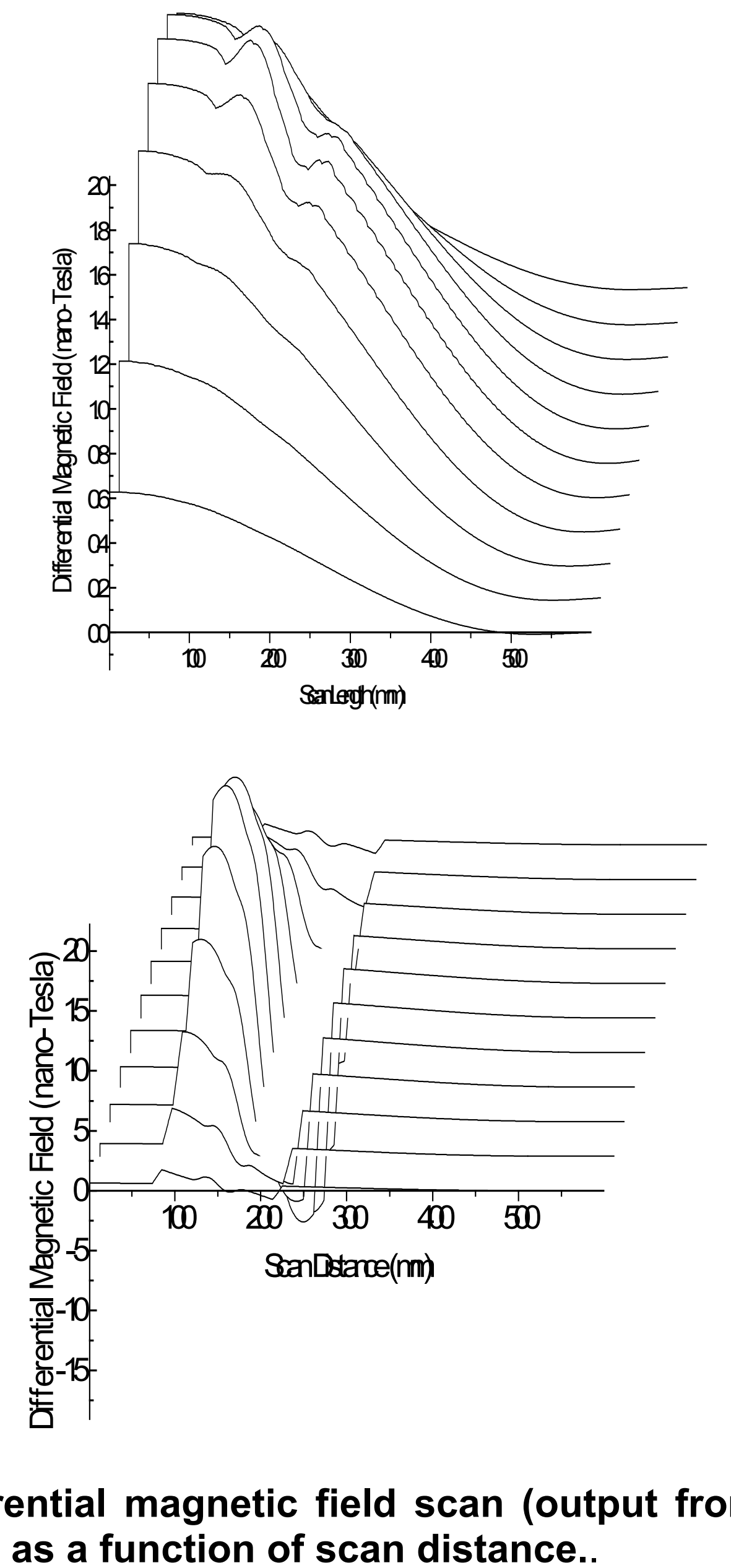

**Figure 9: Differential magnetic field scan (output from first order pickup coil subtraction) as a function of scan distance..**

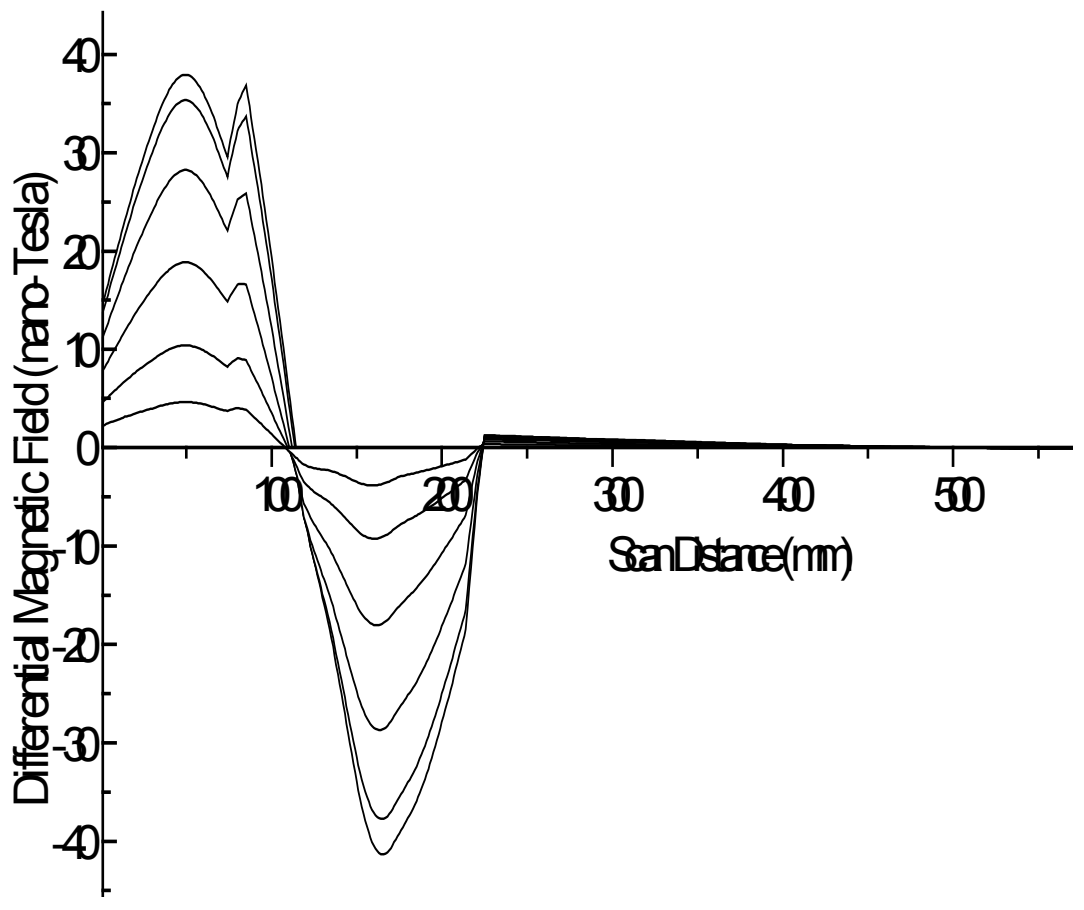

**Figure 10 Line Scan: Tumor at 10cm plus Virtual Organ Background plus Diamagnetic Background**

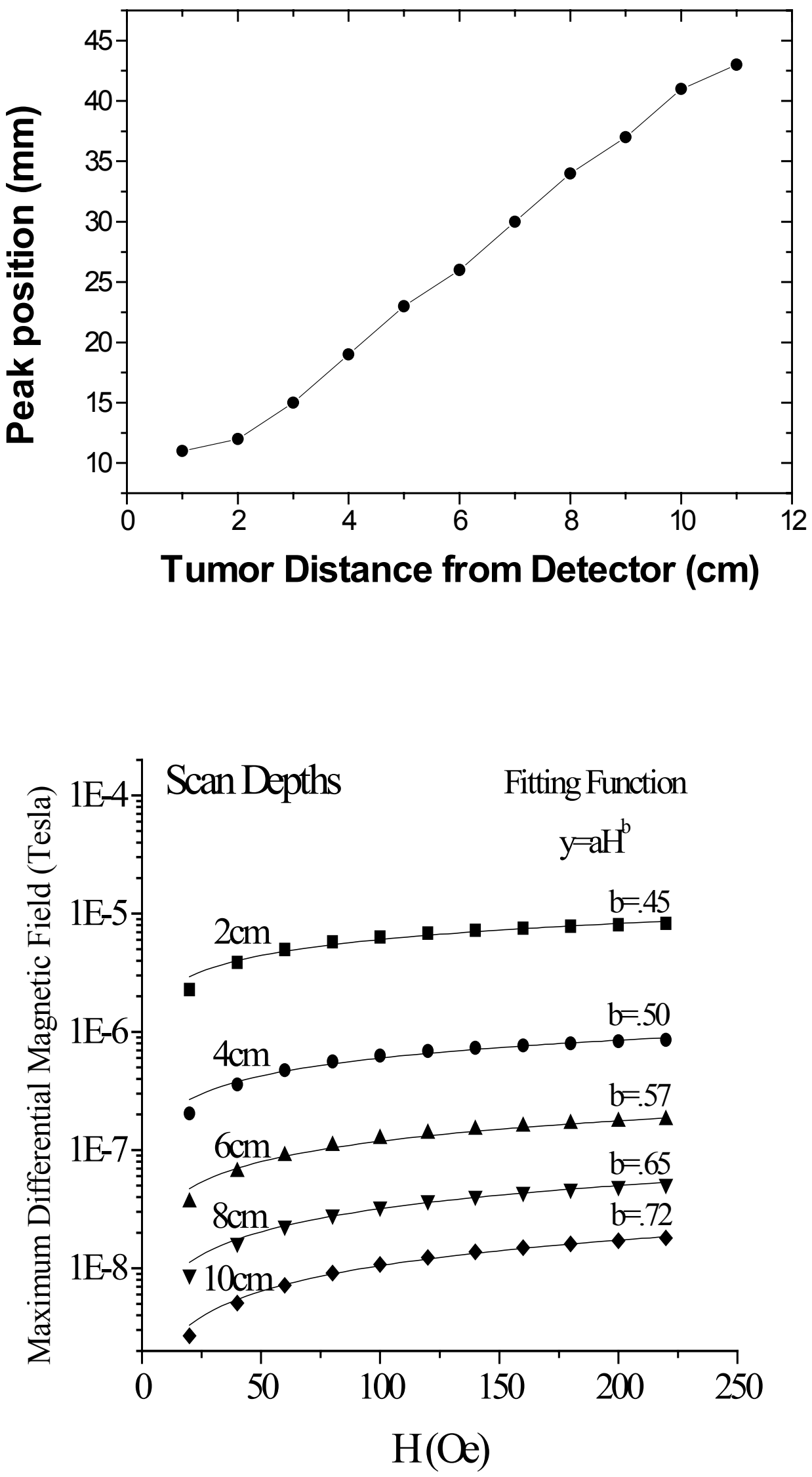

**Figure 11: Depth Dependence of Scanner Signals.**